\documentclass[aps,amsfonts,amssymb]{revtex4}   
 
 \usepackage{graphicx}
 \usepackage{bm}

\begin{document}

\hfill  
\parbox{10cm}{ {KEK-Cosmo-34  KEK-TH-1340}\par}

\title{
Uniqueness theorem for charged dipole rings \\in 
five-dimensional minimal supergravity
}

\vspace{2cm}

\author{Shinya Tomizawa}

\vspace{2cm}
\affiliation{
 KEK Theory Center, Institute of Particle and Nuclear Studies, 
 KEK, Tsukuba, Ibaraki, 305-0801, Japan }

\begin{abstract} 
We show a uniqueness theorem for charged dipole rotating black {\em rings} 
in the bosonic sector of five-dimensional minimal supergravity, 
generalizing our previous work [arXiv:0901.4724] on the uniqueness of charged 
rotating black holes with topologically spherical horizon in the same theory. 
More precisely, assuming the existence of two commuting axial Killing vector 
fields and the same rod structure as the known solutions, 
we prove that an asymptotically flat, stationary charged rotating 
black hole with non-degenerate connected event horizon of cross-section 
topology $S^1\times S^2$ in the five-dimensional Einstein-Maxwell-Chern-Simons 
theory---if exists---is characterized by the mass, charge, two independent 
angular momenta, dipole charge, and the ratio of the $S^2$ radius to 
the $S^1$ radius.  
As anticipated, the necessity of specifying dipole charge---which is not 
a conserved charge---is the new, distinguished ingredient that highlights 
difference between the present theorem and the corresponding theorem 
for vacuum case, as well as difference from the case of topologically 
spherical horizon within the same minimal supergravity. 
We also consider a similar boundary value problem for other topologically 
non-trivial black holes within the same theory, and find that 
generalizing the present uniqueness results to include black lenses%
---provided there exists such a solution in the theory---would not appear 
to be straightforward.

\end{abstract}

\date{\today}

\maketitle

\section{Introduction}
Classifying higher dimensional black holes in supergravity theories is one of 
the key issues toward understanding the structure of string theory. 
In our previous paper \cite{TYI}, we addressed a classification problem of 
black holes in five-dimensional minimal supergravity, and showed that 
if an asymptotically flat, stationary charged rotating black hole solution of 
the theory possesses two rotational symmetries, then it can be uniquely 
specified by its asymptotic conserved charges    
In this version of uniqueness theorem, we restricted attention to the case of 
topologically spherical black holes since in that case, 
relevant boundary value analysis becomes simple and also there is a known 
exact solution \cite{CY96} which appears to be most general as a spherical 
black hole in the five-dimensional minimal supergravity. 
However, topology theorem \cite{Cai,Helfgott,galloway} itself does not 
stop us from considering topologically non-spherical black holes 
as far as horizon cross-section is of positive Yamabe type. 
In fact, a number of topologically non-trivial exact solutions, such as 
black rings and their multiple combinations, have been discovered 
in various theories~\cite{Emparan:2001wn,Pom,MishimaIguchi,diring,saturn,Izumi,bi,EEF,Elvang,EEMR,Y1,Y2,Y3,Yaza08}
It is therefore of considerable interest to consider a generalization of 
our uniqueness result \cite{TYI} to include non-spherical black holes within 
the same supergravity theory. 
The main purpose of this paper is to show, on the basis of Paper~\cite{TYI}, 
a uniqueness theorem for black {\em ring} solutions---assuming their 
existence---in the bosonic sector of five-dimensional minimal 
supergravity theory, or equivalently five-dimensional 
Einstein-Maxwell-Chern-Simons (EMCS) theory with the Chern-Simons coupling 
appropriately chosen. 

\medskip 
Under the assumption that stationary black hole solutions admit 
additionally two independent rotational symmetries 
one can reduce the five-dimensional minimal supergravity theory to precisely 
the same type of non-linear sigma model considered in our previous 
paper~\cite{TYI}, irrespective to the horizon topology. 
One can then construct formally the same divergence identity for 
the sigma model fields on two-dimensional base space. 
Therefore, as briefly discussed in the summary section of Paper~\cite{TYI}, 
the only difference in uniqueness properties between spherical and 
non-spherical black holes should arise in the boundary value 
analysis on the non-linear sigma model. 
The necessary boundary data are given at infinity and at a one-dimensional 
boundary component that corresponds to points of either 
the horizon or ``axis'' of rotational symmetries. 
The latter boundary component is further divided, in a certain manner, 
into a set of segments or intervals of invariant finite (or semi-infinite) 
length. Associated with each interval is an integer-valued 
vector that tells which (or what combination) of the two rotational 
Killing fields vanishes on the interval. 
The collection of such intervals and vectors are called the 
{\em rod-structure} \cite{Harmark} 
(see also \cite{Hollands,HY08}), which in particular specifies the 
horizon topology. 
For example, as discussed in Paper~\cite{TYI}, the rod-structure for 
a single black ring may be given by the following: 
(i) the semi-infinite interval $[c,\infty]$ with the vector $(0,0,1)$, 
(ii) the finite interval $[ck^2,c]$ with $(0,1,0)$, 
(iii) $[-ck^2,ck^2]$ with no vector, corresponding to the event horizon, 
and 
(iv) $[-\infty,-ck^2]$ with $(0,1,0)$, 
where $c>0, \, 0<k^2<1$. We will discuss this in more detail 
in the next section. 
Then, noting that information about horizon topology can be encoded 
in the rod-structure, one might expect that the desired generalization 
of the uniqueness theorem to include non-spherical black holes would be 
straightforwardly achieved by merely specifying appropriate rod-structure 
as well as all possible global conserved charges.  
This is indeed the case for the vacuum solutions \cite{Hollands,MTY}. 
However, one has to be more careful when gauge fields are involved: 
For example, when a rotating black ring couples to Maxwell field, it generates 
type of a dipole field. Accordingly, the {\em dipole charge}---which is not 
a conserved charge---comes to play a role as an additional parameter to 
characterize the solution, as stated already in the first example of 
dipole ring solutions found by Emparan~\cite{Emparan_di}, which are 
electrically coupled to a two form or a dual magnetic one-form field.  
Further examples of dipole rings have been constructed by Elvang 
{\it et al.}~\cite{EEF} in five-dimensional minimal supergravity, starting 
from a seven-parameter family of non-supersymmetric black ring solutions. 
Their solution, however, does not have any limit to a supersymmetric solution, 
and moreover the dipole charge of their solution is not an independent 
parameter: it can be determined in terms of the other asymptotic conserved 
charges. Hence, as conjectured by the authors of \cite{EEF} themselves, 
it is natural to anticipate that there exists a more general non-BPS black 
ring solution characterized by its mass, two independent angular momenta, 
electric charge, and a dipole charge that is independent of the other 
asymptotic conserved charges~\footnote{ 
See also~\cite{Y1,Y2,Y3,Yaza08} about a dipole ring solution or a black Saturn 
solution with a dipole charge in five-dimensional Einstein-Maxwell theory.  
}. 
Although such a seemingly most general dipole ring solution has not been 
discovered yet, assuming its existence, we would like to show the following 
theorem: 

\medskip
\noindent
{\bf Theorem.} 
{\em  
Consider the bosonic part of five-dimensional minimal supergravity, 
i.e., five-dimensional Einstein-Maxwell-Chern-Simons theory with certain 
value of the Chern-Simons coupling, and suppose there exists a regular 
stationary charged rotating black ring with finite temperature: 
that is, a stationary black hole solution that possesses a non-degenerate 
connected event horizon with cross-section topology $S^1 \times S^2$, 
and is regular on and outside the horizon and asymptotically flat 
in the standard sense with spherical spatial infinity.   
If such a black ring solution further admits 
(1) two mutually commuting axial Killing vector fields, in addition to the 
stationary Killing vector field, so that the isometry group is 
${\Bbb R}\times U(1)\times U(1)$, 
and 
(2) the rod-structure of the type (i)--(iv) above, 
then the solution is uniquely characterized by its mass, 
electric charge, two independent angular momenta, dipole charge,  
and the rod data (which corresponds to the ratio of the $S^2$ radius 
to the $S^1$ radius). 
}

\medskip 
Some remarks are in order.  
As discussed in detail in \cite{HY08}, the rod-structure 
(more precisely, the {\em interval structure} of \cite{Hollands,HY08}) 
can specify not only the horizon topology but also topology of the black 
hole exterior region, as well as the action of the rotational symmetries.  
In the present case, restricting the rod-structure as above (i)--(iv), 
the black hole exterior is topologically 
${\Bbb R} \times {\Bbb R}^4 \setminus \{ D^2\times S^2 \} $. 
In fact, all known black ring solutions with a single horizon component 
admit the rod-structure above. 
However, it is not obvious whether any black ring solution must always have 
the rod-structure of this simple type. Also when one wishes to generalize 
the present theorem to include other non-trivial black objects, 
one would need to address the case with more general rod-structure, 
as in fact we will attempt to do so in Sec.~\ref{sec:general}. 
In this regard, a similar uniqueness problem with general rod-structure, 
treating both black rings and spherical holes in a unified manner, 
has been addressed in a rather restricted class of five-dimensional 
Einstein-Maxwell system~\cite{HY}. There, the necessity of 
specifying a dipole charge and other extra charges (for general 
rod-structure case) has also been pointed out.  

\medskip 
In the next section, we prove the above theorem, starting from a brief 
description of general strategy for the black hole uniqueness proof. 
In Sec.~\ref{sec:general}, we study the boundary conditions for black holes 
with other horizon topologies, i.e., black lenses. 
In Sec.~\ref{sec:summary}, we summarize our results and comment on some 
open issues on uniqueness for black lenses and multi-rings.

\section{Proof} 
\label{sec:metric}

Our proof consists of the following three steps (i)--(iii), 
employing the basic techniques of the classic uniqueness proof for 
four-dimensional black holes (see e.g., Ref.~\cite{Heusler} and 
references therein), 
as well as imposing additional conditions upon topology and symmetries. 
In fact, essentially the same strategy has often been used in proof of 
uniqueness theorems proposed for some restricted classes of 
higher dimensional black holes in various---but different from 
the present---context (See e.g., Refs.~\cite{shiromizu,MI,Hollands,MTY,HY08,HY,R,R2,R3,R4,R5,R6,R7,TYI}). 
It goes roughly as follows: 
(i) First, using symmetry conditions, we reduce the theory of interest 
to a certain non-linear sigma model on a two-dimensional base space, $\Sigma$. 
Thanks to the symmetry, $G$, of the sigma model, the set of sigma-model fields,
 $\Phi^A$, on $\Sigma$ can collectively be described in terms of a symmetric, 
unimodular matrix, $M$, on the coset space $G/H$, where $H$ is an isotropy 
subgroup of $G$. Thus, in principle, the solutions of the system can compactly 
be expressed by the matrix $M$. Furthermore, the matrix $M$ formally defines 
a conserved current, $J$, for the solution. 
(ii) Next, we introduce the {\em deviation matrix}, $\Psi$, which is 
essentially the difference between two coset matrices, 
say $M_{[0]}$ and $M_{[1]}$, so that when two solutions coincide with 
each other, the deviation matrix vanishes, and vice versa. 
What we wish to show is that $\Psi$ vanishes over the entire $\Sigma$ 
when two solutions satisfy the same boundary conditions that specify relevant 
physical parameters characterizing the black hole solution of interest. 
For this purpose, we construct a global identity, called the 
{\em Mazur identity}, (the integral version of) which equates an integration 
along the boundary $\partial \Sigma$ of a derivative of the trace of $\Psi$ 
to an integration over the whole base space $\Sigma$ of the trace of 
`square' of the deviation, ${\cal M}$, of the two conserved currents, 
$J_{[0]}$ and $J_{[1]}$. The latter is therefore non-negative. 
The identity is essentially a generalization of the Green's divergence 
identity for the standard Laplace equation. 
(iii) Then, we perform boundary value analysis of the matrix $\Psi$. 
We identify boundary conditions for $M$ that define physical parameters 
characterizing black hole solutions and that guarantee the regularity 
of the solutions. 
Then we examine the behavior of $\Psi$ near $\partial \Sigma$. 
For higher dimensional case, this is the point where the topology and symmetry 
properties, translated into the language of the rod-structure, come to play 
a role as additional parameters to specify solutions. 
When the integral along the boundary $\partial \Sigma$, say the left-side 
of the Mazur identity, vanishes under our boundary conditions, it then 
follows from the right-side of the identity, i.e., the non-negative 
integration over $\Sigma$, that ${\cal M}$ has to vanish, hence 
the two currents, $J_{[0]}$ and $J_{[1]}$, must coincide with each other 
over $\Sigma$, 
implying that the deviation matrix $\Psi$ must be constant over $\Sigma$. 
Then, if $\Psi$ is shown to be zero on some part of the boundary 
$\partial \Sigma$, it follows that $\Psi$ must be identically zero 
over the entire $\Sigma$, thus proving the two solutions, 
$M_{[0]}$ and $M_{[1]}$, must be identical. 

\medskip 
In our present case, the first two steps (i)-(ii) completely parallel 
those in Paper \cite{TYI}, and Step (iii) is the new result of this paper. 
In order to highlight difference from the spherical horizon case and also 
to avoid unnecessary repetition, in the following we provide only some key 
formulas of Steps (i) and (ii), needed in Step (iii) later on, quoting 
from Paper~\cite{TYI}.  

\medskip 
Our starting point is the following five-dimensional 
minimal supergravity action 
\begin{eqnarray}
S=\frac{1}{16\pi}
  \left[ 
        \int d^5x\sqrt{-g}\left(R-\frac{1}{4}F^2\right) 
       -\frac{1}{3\sqrt{3}} \int F\wedge F\wedge A 
  \right] \,, 
\label{action} 
\end{eqnarray} 
where we set the Newton constant to be unity and $F=dA$ with $A$ being 
the gauge potential. Varying this action (\ref{action}), we derive 
the Einstein equations with the standard stress-energy tensor for 
five-dimensional Maxwell field, as well as Maxwell's equations which have 
the extra term coming from the Chern-Simons term of (\ref{action}). 
We are concerned with asymptotically flat, stationary, charged rotating 
black ring solutions of this theory. We additionally impose two independent 
axial symmetries, so that the total isometry group is 
${\Bbb R}\times U(1) \times U(1)$ with ${\Bbb R}$ being stationary 
symmetry, generated by mutually commuting three Killing vector fields 
$\xi_t = \partial /\partial t$ and $\xi_a= (\xi_\phi, \xi_\psi)
=(\partial/\partial \phi, \partial/\partial \psi)$\footnote{
This assumption concerning {\it two} independent axial symmetries is 
not fully justified, as the rigidity theorem~\cite{HIW,MI08,HI} 
guarantees the existence of only a single axial symmetry for 
stationary black holes.  
}. 
Using the Einstein equations and the Maxwell equations, we can show that 
the generators $\xi_t ,\: \xi_a$ of the isometry group satisfy 
type of integrability conditions discussed in Ref.~\cite{Harmark,weyl}. 
As a result, we obtain the coordinate system, $\{t,\phi,\psi,\rho,z\}$, 
in which the metric takes the {\em Weyl-Papapetrou form} 
\begin{eqnarray}
ds^2&=&\lambda_{\phi\phi}(d\phi+a^\phi{}dt)^2
     + \lambda_{\psi\psi}(d\psi+a^\psi{}dt)^2 
\nonumber\\ 
   &&+2\lambda_{\phi\psi}(d\phi+a^\phi{}dt)(d\psi+a^\psi{}dt)
     +|\tau|^{-1}[e^{2\sigma}(d\rho^2+dz^2)-\rho^2 dt^2] \,, 
\end{eqnarray}
and the gauge potential is written, 
\begin{eqnarray} 
 A = \sqrt{3}\psi_a dx^a + A_t dt \,, 
\label{pote:gauge}
\end{eqnarray}
where the coordinates $x^a=(\phi,\psi)$ denote the Killing parameters, 
and thus all functions $\lambda_{ab}$, 
$\tau:=-{\rm det}(\lambda_{ab})$, $a^a$, $\sigma$, and 
$(\psi_a,A_t)$ are independent of $t$ and $x^a$, 
and where the potentials $\psi_a$ are related to Maxwell field 
by eq.~(8) of Paper \cite{TYI} 
[see also Appendix A of Paper~\cite{TYI} for the gauge choice employed 
in eq.~(\ref{pote:gauge})]. 
Note that the coordinates $(\rho,z)$ that span a two-dimensional 
{\em base space}, $\Sigma=\{(\rho,z)|\rho\ge 0,\ -\infty<z<\infty \}$, 
are globally well-defined, harmonic, and mutually conjugate on $\Sigma$. 
See e.g., \cite{Chru08}. 
Furthermore, by using the Maxwell's equation and Einstein's equations, 
we introduce the magnetic potential $\mu$ and twist potentials $\omega_a$ by   
\begin{eqnarray} 
d\mu&=&\frac{1}{\sqrt{3}}*(\xi_\phi\wedge \xi_\psi\wedge F) 
    - \epsilon^{ab}\psi_ad\psi_b \,, 
\label{eq:mu} 
\\
 d\omega_a&=&*(\xi_\phi\wedge \xi_\psi \wedge d\xi_a)+\psi_a(3d\mu+\epsilon^{bc}\psi_bd\psi_c)\,, 
\label{eq:twistpotential} 
\end{eqnarray} 
where $\epsilon^{\phi\psi}=-\epsilon^{\psi\phi}=1$.
Then, the nonlinear sigma-model reduced from the theory (\ref{action}) with 
the symmetry assumptions consists of the target space with the isometry 
$G=G_{2(2)}$ and the eight scalar fields 
$\Phi^A=(\lambda_{ab},\omega_a,\psi_a, \mu)$ on the base space $\Sigma$. 
All the other fields such as $\sigma,a^a$, etc can be determined by $\Phi^A$ 
through the equations of motion. 
It turns out that the sigma model fields, $\Phi^A$, can be expressed 
by a $7 \times 7$ symmetric unimodular coset $G_{2(2)}/SO(4)$ matrix $M$ 
[see eq.~(34) of Paper~\cite{TYI}], as shown by~\cite{MO,MO2,BCCGSW}. 
Then we define the deviation matrix, $\Psi$, for two solutions, 
$M_{[0]}$ and $M_{[1]}$, as in eq.~(42) of Paper \cite{TYI}, 
and derive the Mazur identity, 
\begin{eqnarray}
 \int_{\partial \Sigma}\rho \partial_a {\rm tr} \Psi dS^a 
  = \int_{\Sigma} {\rm tr}({\cal M}^{T}\! \cdot \!{\cal M})\rho d\rho dz \,, 
 \label{eq:id}
\end{eqnarray} 
where {\it dot} denotes the inner product on $\Sigma$. As briefly mentioned 
above, $\cal M$, in the right-side essentially describes the difference 
between two matrix currents $J_{[0]},\:J_{[1]}$, given by eq.~(47) 
of Paper~\cite{TYI}, of which detail is irrelevant to discussion below. 
Our task is to show that the left-side of eq.~(\ref{eq:id}) vanishes on 
the boundary, $\partial \Sigma$, and then show $\Psi$ itself 
vanishes on some part of the boundary.

\medskip 
Now we proceed Step (iii): The boundary value analysis. 
In the Weyl-Papapetrou coordinate system, the boundaries for black rings 
can be described as follows:  
\begin{enumerate}
\item[(i)]
$\psi$-invariant plane: 
$\partial \Sigma_\psi=\{(\rho,z)|\rho=0,c<z<\infty \}$ with the rod vector 
$v=(0,0,1)$ \,, 
\item[(ii)]
$\phi$-invariant plane inside the black ring: 
$\partial \Sigma_{in}=\{(\rho,z)|\rho=0,ck^2<z<c \}$ with the rod vector 
$v=(0,1,0)$ \,,
\item[(iii)] Horizon: 
$\partial \Sigma_{\cal H}=\{(\rho,z)|\ \rho=0,-ck^2<z<ck^2\}$ \,, 
\item[(iv)] 
$\phi$-invariant plane outside the black ring: 
$\partial \Sigma_\phi=\{(\rho,z)|\rho=0,-\infty<z<-ck^2\}$ 
with the rod vector $v=(0,1,0)$\,, 
\item[(v)] Infinity:  
$\partial \Sigma_\infty 
= \{(\rho,z)|\sqrt{\rho^2+z^2}\to 
\infty\ {\rm with}\ z/\sqrt{\rho^2+z^2}\ {\rm finite}  \}$ \,, 
\end{enumerate} 
where the two constants $c$ and $k$ satisfy $c>0$ and $0<k^2<1$.

\medskip 
Therefore, the boundary integral in the left-hand side of the Mazur 
identity, eq.~(\ref{eq:id}), is decomposed into the integrals over the four
rods (i)--(iv), and the integral at infinity (v), as 
\begin{eqnarray}
\int_{\partial \Sigma}\rho\partial_p{\rm tr}\Psi dS^p 
&=& \int_{-\infty}^{-ck^2}\rho\frac{\partial {\rm tr}\Psi }{\partial z}dz 
   +\int_{-ck^2}^{ck^2}\rho\frac{\partial {\rm tr}\Psi }{\partial z}dz 
+\int_{ck^2}^{c}\rho\frac{\partial {\rm tr}\Psi }{\partial z}dz 
\nonumber\\
 && +\int_{c}^{\infty}\rho\frac{\partial {\rm tr}\Psi }{\partial z}dz 
    +\int_{\partial\Sigma_\infty}\rho\partial_a{\rm tr}\Psi dS^a \,.   
\label{eq:descom}
\label{eq:integral} 
\end{eqnarray}
Note that the only difference between black holes and black rings 
appears at the third term in the right-side of eq.~(\ref{eq:descom}), 
which corresponds to the integral over the $\phi$-invariant plane inside 
the black ring. As will be seen below, because of the existence of this third 
integral, a dipole charge comes to appear in our boundary conditions.

\medskip 
We examine the behavior of $\Psi$ at each boundary, (i)--(v), 
separately, starting from analysis at Infinity~(v).  

\medskip 
\noindent
(v) Infinity: 
$\partial\Sigma_\infty=\{(\rho,z)|\ \sqrt{\rho^2+z^2}\to\infty$ with 
$z/\sqrt{\rho^2+z^2}$ kept finite $\}$. 
Since we are concerned with asymptotically flat solutions in 
the standard sense and the behavior of the scalar fields near infinity 
does not depend on what the topology of the horizon is, 
the discussion here is the same as the case of a spherical horizon 
topology~\cite{TYI}. So we have 
\begin{eqnarray}
 \rho\ \partial_p {\rm tr} \Psi dS^p 
 \simeq {\cal O}\left(\frac{1}{\sqrt{\rho^2+z^2}}\right) \,.
\end{eqnarray}

\medskip
\noindent
(i) $\psi$-invariant plane: 
$\partial\Sigma_\psi=\{(\rho,z)|\rho=0,\ c<z<\infty\}$. 
This part is essentially the same as the boundary analysis at 
$\phi$-invariant plane of Paper \cite{TYI}. Note that in this paper, 
we are taking $\psi$ as the Killing parameter along the $S^1$ sector of 
the ring solution, and for this reason the role of $\partial \Sigma_\psi$ 
is played by the `$\phi$-invariant plane' of Paper \cite{TYI}. 
Therefore the behavior of $\Psi$ near $\partial \Sigma_\psi$ 
can be read off from the formulas of eqs.~(63)--(70) of Paper~\cite{TYI}. 
As a result, for two solutions, $M_{[0]}$ and $M_{[1]}$, with the same mass, 
the same angular momenta, and the same electric charge, we have, 
\begin{eqnarray}
\rho\ \partial_z{\rm tr}\ \Psi\simeq O(\rho) \,. 
\end{eqnarray}

\medskip 
\noindent 
(iv) $\phi$-invariant plane outside the black ring: 
$\partial\Sigma_\phi=\{(\rho,z)|\rho=0,\ -\infty<z<-ck^2\}$. 
Similarly, the behavior of $\Psi$ around $\partial\Sigma_\phi$ can be 
read off from eqs.~(75)--(82) of Paper~\cite{TYI}, and 
we have 
$
\rho\ \partial_z{\rm tr}\ \Psi \simeq O(\rho) 
$, for $\rho \to 0$. 

\medskip 
\noindent 
(iii) Horizon: $\partial \Sigma_{\cal H}=\{(\rho,z)|\ \rho=0, \ -ck^2<z<ck^2 \}$.
The regularity on the horizon requires that 
for $\rho \to 0$,   
\begin{eqnarray}
&&\lambda_{ab}\simeq {\cal O}(1),\quad \omega_{a}\simeq {\cal O}(1) \,,\psi_{a}\simeq {\cal O}(1), \quad \mu\simeq {\cal O}(1) \,. 
\end{eqnarray}
Thus, we have for $\rho \to 0$, 
$
\rho\ \partial_z{\rm tr}\ \Psi\simeq O(\rho) 
$.

\medskip 
\noindent 
(ii) $\phi$-invariant plane inside the black ring: 
$\partial\Sigma_{in}=\{(\rho,z)|\rho=0,\ ck^2<z<c\}$. 
This is the key part of the present boundary value analysis. 
As in the case~(iv), the regularity requires that the potentials, 
$\lambda_{ab}$, must behave as 
\begin{eqnarray}
&&\lambda_{\phi\phi}\simeq {\cal O}(\rho^2),\label{eq:d1}\\
&&\lambda_{\phi\psi}\simeq  {\cal O}(\rho^2),\label{eq:d2}\\
&&\lambda_{\psi\psi}\simeq  {\cal O}(1).\label{eq:d3}
\end{eqnarray}
The dipole charge for a black ring is defined by 
\begin{eqnarray}
q&:=&\frac{1}{2\pi}\int_{S^2}F \nonumber \\
 &=&\frac{1}{2\pi}\int_{S^2}A_{\phi,z}dz\wedge d\phi \nonumber \\
 &=&\sqrt{3}\left[\psi_\phi(\rho=0,z=ck^2)-\psi_\phi(\rho=0,z=-ck^2)\right]
\nonumber \\
 &=&\sqrt{3}\psi_\phi(\rho=0,z=ck^2) \,, 
\label{eq:dipole} 
\end{eqnarray}
where $S^2$ denotes the two-sphere spatial cross section of the black ring 
horizon and where, at the fourth equality, we have used the fact that 
$\psi_\phi \simeq {\cal O}(\rho^2)$ on the $\phi$-invariant plane 
$\partial \Sigma_\phi$ [eq.~(81) of Paper~\cite{TYI}]. 
Note that the derivative of the electric potential, $d\psi_\phi$, vanishes 
on $\partial\Sigma_{in}$ by definition, and hence  $\psi_\phi$ is constant 
over $\partial\Sigma_{in}$. Therefore, from eq.~(\ref{eq:dipole}), 
we immediately find that the electric potential, $\psi_\phi$, must behave as 
\begin{eqnarray}
\psi_\phi\simeq \frac{q}{\sqrt{3}}+{\cal O}(\rho^2)\label{eq:d7} 
\end{eqnarray}
in a neighborhood of $\partial\Sigma_{in}$. 
We cannot determine how the other magnetic potential, $\psi_\psi$, behaves on 
$\partial \Sigma_{in}$, and therefore we write 
\begin{eqnarray}
\psi_\psi\simeq f(z)+{\cal O}(\rho^2) \,, 
\label{eq:d6}  
\end{eqnarray}
where $f(z)$ is some function depending only on $z$.

\medskip
Next, we consider the magnetic potential $\mu$ on the $\phi$-invariant plane 
inside the black ring $\partial \Sigma_{in}$. 
The magnetic potential, $\mu$, satisfies, eq.~(\ref{eq:mu}), i.e.,  
\begin{eqnarray}
 d\mu = \frac{1}{\sqrt{3}}*(\xi_\phi\wedge \xi_\psi\wedge F) 
      - (\psi_\phi d\psi_\psi -\psi_\psi d\psi_\phi) \,.
\end{eqnarray}
The first term vanishes on $\partial \Sigma_{in}$ by definition. 
Substituting eq.~(\ref{eq:d7}) into the above equation, 
we find that the derivative of $\mu$ can be written as 
\begin{eqnarray}
d\mu =-\frac{q}{\sqrt{3}}d\psi_\psi \,. 
\label{eq:dmu} 
\end{eqnarray} 
Hence, integrating (\ref{eq:dmu}), we obtain
\begin{eqnarray}
\mu=-\frac{q}{\sqrt{3}}\psi_\psi+c_{in},
\end{eqnarray}
where $c_{in}$ is a constant. 
Here, we note from the analysis (i) on $\partial \Sigma_\psi$ that 
$\psi_\psi \rightarrow 0$ [eq.~(68) of Paper~\cite{TYI}] and 
$\mu \rightarrow {2Q}/{\sqrt{3}\pi}$ [eq.~(70) of Paper~\cite{TYI}] 
at the center of the black ring, $\rho=0,z=c$. 
Then, the constant is determined as 
\begin{eqnarray}
c_{in}=\frac{2Q}{\sqrt{3}\pi}.
\end{eqnarray}
Thus, in terms of the undetermined function, $f(z)$, of $z$, the electric 
charge, $Q$, and the dipole charge, $q$, we can obtain the behavior of 
the magnetic potential, $\mu$, near $\partial \Sigma_{in}$ as follows. 
\begin{eqnarray}
\mu\simeq -\frac{q}{\sqrt{3}}f(z)+\frac{2Q}{\sqrt{3}\pi}+{\cal O}(\rho^2).\label{eq:d8}
\end{eqnarray}

\medskip
Furthermore, consider the boundary condition for the twist potentials, 
$\omega_a$, on $\partial\Sigma_{in}$. Recall that they are given by 
\begin{eqnarray}
d\omega_a=*(\xi_\phi\wedge \xi_\psi\wedge d\xi_a) 
         +\psi_a(3d\mu+\psi_\phi d\psi_\psi -\psi_\psi d\psi_\phi) \,.
\end{eqnarray}
The first term vanishes on the $\psi$-invariant plane, $\partial\Sigma_{in}$, 
by definition. Substituting eqs.~(\ref{eq:d7}) and (\ref{eq:d8}) into 
the above equation, we find that the derivative of the twist potentials, 
$\omega_a$, can be written as  
\begin{eqnarray}
d\omega_a=-\frac{2q}{\sqrt{3}}\psi_ad\psi_\psi.
\end{eqnarray} 
Then, by using eqs.~(\ref{eq:d7})-(\ref{eq:d6}), they can be rewritten as 
\begin{eqnarray}
d\omega_\phi= -\frac{2}{3}q^2df(z),\quad d\omega_\psi 
            = -\frac{2}{\sqrt{3}}q f(z) df(z) \,. 
\label{eq:domega}
\end{eqnarray}
Integrating eq.~(\ref{eq:domega}) on $\partial\Sigma_{in}$, we obtain 
\begin{eqnarray}
\omega_\phi= - \frac{2}{3}q^2 f(z)+c_\phi, \quad \omega_\psi 
           = - \frac{1}{\sqrt{3}}q f(z)^2+c_\psi \,, 
\end{eqnarray}
where $c_\phi$ and $c_\psi$ are arbitrary constants. 
From the analysis (i) on $\partial \Sigma_\psi$ we have 
\begin{eqnarray}
\omega_a=-\frac{2J_a}{\pi},\quad \psi_\psi=0
\end{eqnarray}
at the center of the black ring, $\rho=0, z=c$ 
[eqs.~(66)--(68) of Paper~\cite{TYI}].  
Hence, the constants, $c_\phi$ and $c_\psi$, can be determined as
\begin{eqnarray}
c_\phi=-\frac{2J_\phi}{\pi},\quad c_\psi=-\frac{2J_\psi}{\pi}.
\end{eqnarray}
As a result, we find that the two twist potentials $\omega_\phi$ and $\omega_\psi$ must behave as 
\begin{eqnarray}
&&\omega_\phi\simeq -\frac{2}{3}q^2 f(z)-\frac{2J_\phi}{\pi}+{\cal O}(\rho^2),\label{eq:d4}\\
&& \omega_\psi\simeq-\frac{1}{\sqrt{3}}q f(z)^2-\frac{2J_\psi}{\pi}+{\cal O}(\rho^2)\label{eq:d5},
\end{eqnarray}
on $\partial\Sigma_{in}$.
Thus, from eqs.~(\ref{eq:d1})-(\ref{eq:d3}), (\ref{eq:d7})-(\ref{eq:d6}), 
(\ref{eq:d8}), and (\ref{eq:d4})-(\ref{eq:d5}), 
we find for $\rho \to 0$, 
$\rho\ \partial_z{\rm tr}\ \Psi\simeq O(\rho) $. 
We emphasize here that in order to obtain this result, we do not need to, 
in advance, specify the functions, $f(z)_{[0]}, \: f(z)_{[1]}$, 
in the two solutions.

\medskip 
We conclude from (i)--(v) that the boundary integral vanishes on each rod 
and the infinity. 
We can also find, by continuity, that the boundary integral is bounded, 
hence vanishing, at the points where two adjacent rods meet.   
The deviation matrix, $\Psi$, is constant and has the asymptotic behavior, 
$\Psi\to 0$. Therefore, $\Psi$ vanishes over $\Sigma$, and the two 
configurations, $M_{[0]}$ and $M_{[1]}$, coincide with each other for the two 
black ring solutions with the same mass, angular momenta, electric charge, 
dipole charge and rod structure (i.e., $k^2$). 
This completes our proof for the uniqueness theorem.

\section{Boundary value problem for black lens}
\label{sec:general}

As discussed in~\cite{Hollands}, under the existence of two commuting axial 
Killing vectors, the cross-section topology of each connected 
component of the event horizon of stationary vacuum black hole solutions 
must be either $S^3$, $S^1\times S^2$ or a lens space. 
In this section we would like to consider the boundary value problem for 
an asymptotically flat, black lens, though such a solution has not been found 
even in the vacuum case. The rod structure for a black lens was given 
by Evslin~\cite{Evslin}. 
In the Weyl-Papapetrou coordinate system, the boundaries for a black lens 
with the $L(p;1)$ horizon topology, if exists, can be given as follows:  
\begin{enumerate}
\item[(i)]
$\psi$-invariant plane: 
$\partial \Sigma_\psi=\{(\rho,z)|\rho=0,c<z<\infty \}$ with the rod vector 
$v=(0,0,1)$ \,, 
\item[(ii)]
Inner axis 
$\partial \Sigma_{in}=\{(\rho,z)|\rho=0,ck^2<z<c \}$ with the rod vector 
$v=(0,1,p)$ \,,
\item[(iii)] Horizon: 
$\partial \Sigma_{\cal H}=\{(\rho,z)|\ \rho=0,-ck^2<z<ck^2\}$ \,, 
\item[(iv)] 
$\phi$-invariant plane
$\partial \Sigma_\phi=\{(\rho,z)|\rho=0,-\infty<z<-ck^2\}$ 
with the rod vector $v=(0,1,0)$ \,, 
\item[(v)] Infinity:  
$\partial \Sigma_\infty 
= \{(\rho,z)|\sqrt{\rho^2+z^2}\to 
\infty\ {\rm with}\ z/\sqrt{\rho^2+z^2}\ {\rm finite}  \}$ \,, 
\end{enumerate} 
where constants $c$ and $k$ satisfy $c>0$ and $0<k^2<1$. 

\medskip 
For the boundaries ${\rm (i),(iii),(iv)}$ and ${\rm (v)}$, 
the boundary conditions of the scalar fields, $\Phi^A$, are exactly the 
same as those of black rings. Therefore we consider only ${\rm (ii)}$. 
First, note that the rod vector, 
$v=\partial/\partial\phi+p\partial/ \partial \psi$, has fix points for 
$\rho=0,\ z\in[ck^2,c]$, i.e., 
\begin{eqnarray}
g(v,v)=0 \Longleftrightarrow \lambda_{\phi\phi}+2p \lambda_{\phi\psi} +p^2\lambda_{\psi\psi}=0. \label{eq:vv}
\end{eqnarray} 
On the inner axis, we have 
\begin{eqnarray}
\tau=\lambda_{\phi\psi}^2-\lambda_{\phi\phi}\lambda_{\psi\psi}=0.
\end{eqnarray}
Therefore, we find that near the inner axis, the potentials, $\lambda_{ab}$, must behaves as
\begin{eqnarray}
&&\lambda_{\phi\phi}\simeq p^2g(z)+{\cal O}(\rho^2), \\
&&\lambda_{\phi\psi}\simeq -pg(z)+{\cal O}(\rho^2), \\
&&\lambda_{\psi\psi}\simeq g(z)+{\cal O}(\rho^2), 
\end{eqnarray}
where $g(z)$ is some function of $z$. 

\medskip 
Next, consider the boundary conditions for the electric potentials $\psi_a$. 
It follows from eq.~(\ref{eq:vv}) that for $\rho=0,\ z\in[ck^2,c]$, 
\begin{eqnarray}
0=-i_vF=d\psi_\phi+pd\psi_\psi \,. 
\end{eqnarray}
Integrating this, we obtain
\begin{eqnarray}
\psi_\phi+p\psi_\psi=c_0 \,, 
\label{eq:c0}
\end{eqnarray}
where $c_0$ is a constant. 
Therefore we can set the electric potentials to behave as 
\begin{eqnarray}
&&\psi_\phi\simeq c_0-p h(z)+{\cal O}(\rho^2) \,, \label{eq:d6b} \\
&&\psi_\psi \simeq  h(z)+{\cal O}(\rho^2) \,, \label{eq:d7b}
\end{eqnarray}
with $h(z)$ being some function of $ z$. 

\medskip
We further consider the behavior of the magnetic potential $\mu$ defined 
by eq.~(\ref{eq:mu}). Since the norm of the rod vector $v$ vanishes 
over the inner axis, the first term in the right-hand side of 
eq.~(\ref{eq:mu}) vanishes there. 
Then, it follows from eqs.~(\ref{eq:d6b}) and (\ref{eq:d7b}) that 
the derivative of the magnetic potential, $\mu$, is given by 
\begin{eqnarray}
d\mu=-c_0dh(z).
\end{eqnarray} 
Integrating this, we obtain
\begin{eqnarray}
\mu=-c_0 h(z)+c_1 \,, 
\end{eqnarray}
where $c_1$ is an integration constant. 
Here, note that $\mu=2Q/(\sqrt{3}\pi),$ and $\psi_\psi=0\ (i.e., h(z=0)=0)$ 
hold at $\rho=0,\ z=c$. Therefore, the constant $c_1$ is determined as 
\begin{eqnarray}
c_1=\frac{2Q}{\sqrt{3}\pi}.
\end{eqnarray}
Thus, we find that near the inner axis, the magnetic potential, $\mu$, must 
behave as
\begin{eqnarray}
\mu\simeq -c_0 h(z)+\frac{2Q}{\sqrt{3}\pi}+{\cal O}({\rho^2}).
\end{eqnarray}

\medskip
Finally, let us consider the twist potentials $\omega_a$ on the inner axis.
From eqs.~(\ref{eq:d6b}) and (\ref{eq:d7b}), the derivatives of the twist 
potentials on the inner axis are give by 
\begin{eqnarray}
d\omega_a=-2c_0 \psi_a dh(z) \,. 
\end{eqnarray}
Then, it follows that $\omega_a$ can be written in terms of the integration 
constants 
\begin{eqnarray}
\omega_\phi=-2c_0^2h(z)+pc_0 h(z)^2+c_2,\quad \omega_\psi=-c_0h(z)^2+c_3.
\end{eqnarray}
We easily find that
\begin{eqnarray}
\omega_\phi=-\frac{2J_a}{\pi},\quad \psi_\psi =0 \,.
\end{eqnarray}
at $\rho=0,\ z=c$. From continuity of the potentials, 
the constants $c_2$ and $c_3$ can be determined as 
\begin{eqnarray}
c_2=-\frac{2J_\phi}{\pi},\quad c_3=-\frac{2J_\psi}{\pi} \,.
\end{eqnarray}
Therefore, we find the twist potentials behave as 
\begin{eqnarray}
&&\omega_\phi\simeq-2c_0^2h(z)+pc_0 h(z)^2-\frac{2J_\phi}{\pi}
                   +{\cal O}({\rho^2}) \,, 
\\
&&\omega_\psi\simeq -c_0h(z)^2-\frac{2J_\psi}{\pi}+{\cal O}({\rho^2}) \,.
\end{eqnarray}
near the inner axis.

\medskip 
From the above behavior of the scalar fields, we find that the leading term of the boundary 
integral $\int \rho\ \partial_z {\rm tr}\ \Psi dz$ is proportional to 
$(c_{0[0]}-c_{0[1]})\rho^{-3}$. 
Therefore, if the integration constants, 
$c_{0[0]}$ and $c_{0[1]}$, for two solutions with the same mass, two angular 
momenta and electric charge do not coincide with each other, 
the boundary integral does not vanish on the inner axis. 
Since in the vacuum case, the constant $c_0$ vanishes, our 
analysis above immediately implies that the boundary integral vanishes 
on the inner axis. 
This coincides with the results obtained in~\cite{HY}. 
However, in the present case with Maxwell field being non-vanishing, 
there seems to be no obvious way to relate the constant $c_0$ to 
asymptotic charges and the rod data.

\section{Summary and discussion}\label{sec:summary} 
In this paper, we have considered asymptotically flat, stationary charged 
rotating black rings, i.e., holes having $S^1\times S^2$ horizon 
cross-section topology, in the bosonic sector of five-dimensional minimal 
supergravity, and have proven a uniqueness theorem that states that 
under the assumptions of the existence of two commuting axial isometries, 
such a black ring with non-degenerate horizon is characterized by the rod data 
which corresponds to the ratio of the $S^2$ radius to the $S^1$ radius, 
and the following five charges: i.e., the mass, charge, two independent 
angular momenta, and dipole charge. 
As mentioned before, so far no such black ring solutions have been 
discovered. 
The solution obtained by Elvang {\it et al.}~\cite{EEF} admit 
no limit to a supersymmetric black ring solution because the solution does 
not have enough independent parameters, i.e., the dipole charge is not an 
independent parameter, except the case in which the net charge $Q$ vanishes. 
One can expect that there should exist a non-supersymmetric charged dipole 
ring solution with five independent parameters. If it is the case, 
our theorem states that such a solution must be uniquely determined 
by the five charges mentioned above and the rod data. 
Note that such a most general charged dipole ring solution may turn out 
to be generically unbalanced, having a naked conical singular disk 
inside the ring, as in the first example of a static black ring 
in vacuum~\cite{weyl}. Even in the case, our theorem would still apply 
(with removing the requirement that the spacetime itself be regular on 
and outside the horizon in the statement of the theorem), since the existence 
of such a conical singularity does not affect the regularity of our target 
space scalar fields in a neighborhood of the boundary.

\medskip 
We have also considered a similar boundary value problem for asymptotically 
flat, black lens solutions---even though no such a black lens solution 
has been found so far. We have not been able to relate the integration 
constant $c_0$ in eq.~(\ref{eq:c0}) to any of the other charges, except for 
the vacuum case ($Q=q=0$). 
This indicates that the constant $c_0$ arises as a result of interplay 
between the non-vanishing gauge field and non-trivial topology of the horizon, 
just like the dipole charge in the black ring case, and therefore may possibly 
play a role of an independent parameter to uniquely specify a black lens 
solution (if exists) in the minimal supergravity. 
In this paper, however, we have not been able to identify the physical 
interpretation of $c_0$. 
We also expect that a similar problem just mentioned 
above may occur when considering uniqueness theorems for multi-rings, 
black-Saturn, or more complicated black objects which couple to some 
non-vanishing gauge field and which admit the rod-structure that contains 
a rod similar to $\partial \Sigma_{in}$ in the above black-lens example. 
This issue deserves to further study.

\section*{Acknowledgements} 

We would like to thank P. Figueras for valuable discussions and 
comments. S.T. is supported by the JSPS under Contract No. 20-10616. 
The work of Y.Y. is supported by the Grant-in Aid for Scientific 
Research (No. 21244003) from Japan Ministry of Education. 
The work of A.I. is supported by the Grant-in Aid for Scientific Research 
from Japan Ministry of Education.


\begin{thebibliography}{99}

\bibitem{TYI}
S. Tomizawa, Y. Yasui and A. Ishibashi, Phys. Rev. D {\bf 79}, 124023 (2009).


\bibitem{CY96}
M. Cveti\v{c} and D. Youm, Nucl. Phys. B {\bf 476}, 118 (1996).

\bibitem{Cai}
M. I. Cai and G. J. Galloway, Class Quant. Grav. {\bf 18}, 2707 (2001).
\bibitem{Helfgott}
C. Helfgott, Y. Oz and  Y. Yanay, JHEP 0602, 025 (2006).
\bibitem{galloway}
G. J. Galloway and R. Schoen, Commun. Math. Phys. {\bf 266}, 571 (2006). 





\bibitem{Emparan:2001wn}
R.~Emparan and H.~S.~Reall,
Phys.\ Rev.\ Lett.\  {\bf 88}, 101101 (2002).
\bibitem{Pom}
A.A. Pomeransky and R.A. Sen'kov, e-Print: arXiv:hep-th/0612005.
\bibitem{MishimaIguchi}
T.~Mishima and H.~Iguchi, Phys. Rev. D {\bf 73}, 044030 (2006).
\bibitem{diring} 
H. Iguchi and T. Mishima, Phys. Rev. D {\bf 75}, 064018 (2007).
\bibitem{saturn}
H. Elvang and P. Figueras, JHEP {\bf 0705}, 050 (2007).
\bibitem{Izumi}
K. Izumi, Prog. Theor. Phys. {\bf 119}, 757 (2008).
\bibitem{bi}
H. Elvang and M. J. Rodriguez, JHEP {\bf 0804} ,045 (2008).














\bibitem{EEF}
H. Elvang, R. Emparan and P. Figueras, JHEP {\bf 0502}, 031 (2005).
\bibitem{Elvang}
H. Elvang, R. Emparan, D. Mateos and H. S. Reall,
Phys. Rev. Lett. {\bf 93}, 211302 (2004).
\bibitem{EEMR}
H. Elvang, R. Emparan, D. Mateos and H. S. Reall, Phys. Rev. D {\bf 71}, 024033 (2005).
\bibitem{Y1}
S. S. Yazadjiev, Phys. Rev. D {\bf 77}, 127501 (2008). 
\bibitem{Y2}
S. S. Yazadjiev, Phys. Rev. D {\bf 76}, 064011 (2007).
\bibitem{Y3}
S. S. Yazadjiev, Phys. Rev. D {\bf 73}, 104007 (2006).
\bibitem{Yaza08}
S.S.~Yazadjiev, Phys. Rev. D {\bf 78}, 064032 (2008). 
\bibitem{Harmark}
T. Harmark, Phys. Rev. D {\bf 70}, 124002 (2004);
T. Harmark and P.~Olesen, Phys. Rev. D {\bf 72}, 124017 (2005).
\bibitem{Hollands}
S. Hollands and S. Yazadjiev, Commun. Math. Phys. {\bf 283}, 749 (2008).
\bibitem{HY08}  
S. Hollands and S.S.~Yazadjiev, Arxiv: 0812.3036 [gr-qc].
\bibitem{MTY}
Y. Morisawa, S. Tomizawa and Y. Yasui, Phys. Rev. D {\bf 77}, 064019 (2008). 
\bibitem{Emparan_di}
R.~Emparan, JHEP {\bf 03}, 064 (2004).
\bibitem{Heusler}
M. Heusler, {\it Black Hole Uniqueness Theorems} (Cambridge Univ. Press, London, 1996).
\bibitem{shiromizu}
G. W. Gibbons, D. Ida and T. Shiromizu,
Prog. Theor. Phys. Suppl. {\bf 148}, 284 (2002);\\
G. W. Gibbons, D. Ida and T. Shiromizu,
Phys. Rev. Lett. {\bf 89}, 041101 (2002).
\bibitem{MI}
Y.~Morisawa and D.~Ida,
Phys.\ Rev.\ D {\bf 69}, 124005 (2004).
\bibitem{HY}
S. Hollands and S. Yazadjiev, Class. Quantum Grav. {\bf 25}, 095010 (2008).

\bibitem{R}
M. Rogatko, Class. Quant. Grav. {\bf 19}, 875 (2002).
\bibitem{R2}
M. Rogatko, Class. Quant. Grav. {\bf 19}, L151 (2002).
\bibitem{R3}
M. Rogatko, Phys. Rev. D {\bf 67}, 084025 (2003).
\bibitem{R4}
M. Rogatko, Phys. Rev. D {\bf 70}, 044023 (2004).
\bibitem{R5}
M. Rogatko, Phys. Rev. D {\bf 70}, 084025 (2004).
\bibitem{R6}
M. Rogatko, Phys. Rev. D{\bf 73}, 124027 (2006).
\bibitem{R7}
M. Rogatko, Phys. Rev. D {\bf 77}, 124037 (2008).
\bibitem{weyl}  
R. Emparan and H. S. Reall, Phys. Rev. D {\bf 65}, 084025 (2002).
\bibitem{Chru08}
B.~Carter, Phys. Rev. Lett. {\bf 26}, 331 (1971);  
P.T.~Chrusciel, arXive:0812.3424 [gr-qc]. 
\bibitem{BCCGSW}
A. Bouchareb, G. Clement, C-M Chen, D. V. Gal'tsov, N. G. Scherbluk and T. Wolf. Phys. Rev. D {\bf 76},104032 (2007); Erratum-ibid. D {\bf 78}, 029901 (2008).
\bibitem{MO}
S. Mizoguchi and N. Ohta, Phys. Lett. B {\bf 441}, 123 (1998).
\bibitem{MO2}
S. Mizoguchi and G. Schr\"oder, Class. Quan. Grav, {\bf 17}, 835 (2000).
 \bibitem{Evslin}
J, Evslin, JHEP {\bf 0809}, 004 (2008).














































\bibitem{HIW}
S. Hollands, A. Ishibashi and R. M. Wald, 
Commun. Math. Phys. {\bf 271}, 699 (2007).


\bibitem{MI08}
V. Moncrief and J. Isenberg, 
Class. Quan. Grav, {\bf 25}, 195015 (2008). 

\bibitem{HI}
S. Hollands and A. Ishibashi, 
Commun. Math. Phys. {\bf 291}, 403 (2009). 




\end{thebibliography}
\end{document}